\documentclass[%
 reprint,
superscriptaddress,
 amsmath,amssymb,
 aps,
prl,
]{revtex4-1}

\usepackage{graphicx}
\usepackage{dcolumn}
\usepackage{bm}

\usepackage{siunitx}

\newcommand{\m}{\mathrm}

\newcommand{\fref}[1]{Fig.~\ref{#1}}

\begin{document}


\title{Optomechanical measurement of a millimeter-sized mechanical oscillator near the quantum limit}

\author{J. T. Santos}
\affiliation{Department of Applied Physics, Aalto University, P.O. Box 15100, FI-00076 AALTO, Finland}%

\author{J. Li}
\affiliation{School of Engineering, University of Glasgow, Glasgow, G12 8LT, United Kingdom}
\affiliation{Interdisciplinary Center of Quantum Information, College of Science, National University of Defense Technology, Changsha 410073, China}

\author{J. Ilves}
\affiliation{Department of Applied Physics, Aalto University, P.O. Box 15100, FI-00076 AALTO, Finland}%

\author{C. F. Ockeloen-Korppi}
\affiliation{Department of Applied Physics, Aalto University, P.O. Box 15100, FI-00076 AALTO, Finland}%

\author{M. Sillanp\"a\"a}
\affiliation{Department of Applied Physics, Aalto University, P.O. Box 15100, FI-00076 AALTO, Finland}%


\begin{abstract}
\noindent Cavity optomechanics is a tool to study the interaction between light and micromechanical motion. Here we observe near-quantum limited optomechanical physics in a truly macroscopic oscillator. As the mechanical system, we use a \si{\milli\meter}-sized piezoelectric quartz disk oscillator. Its motion is coupled to a charge qubit which translates the piezo-induced charge into an effective radiation-pressure interaction between the disk and a microwave cavity. We measure the thermal motion of the lowest mechanical shear mode at \SI{7}{\mega\hertz} down to \SI{35}{\milli\kelvin}, corresponding to roughly \si{10^2} quanta in a \SI{20}{\milli\gram} oscillator. The work opens up opportunities for macroscopic quantum experiments.
\end{abstract}

\maketitle

Small but nevertheless macroscopic systems have been operated in the limit where they exhibit quantum mechanical behavior in some of their degrees of freedom. One of the most successful systems have been the superconducting quantum bits (qubits) \cite{nakamura1999, vion2002}, where the most complicated man-made quantum states have been constructed \cite{hofheinz2009, vlastakis2013}. Observing mechanical oscillators at the quantum limit of their motion, where the excess phonon number $n_m$ approaches zero, was a long-standing goal. The ground state was reached some years ago \cite{o2010, teufel2011, chan2011}. In earlier experiments,  $n_m<10^2$  was reached with nanostrings already a decade ago  \cite{lahaye2004, etaki2008, anetsberger2009}. In the cavity optomechanical scheme, where optical and mechanical resonances are coupled, micro mirrors up to \SI{0.2}{\milli\gram} weight \cite{arcizet2006, gigan2006, groblacher2009}, cantilevers \cite{metzger2004, kleckner2006, vinante2016} up to \SI{0.5}{\milli\meter} long, or nitride membranes \cite{sankey2010} have been used. Ground-state cooling \cite{teufel2011}, entanglement \cite{palomaki2013}, and squeezed mechanical states \cite{wollman2015, pirkkalainen2015, lecocq2015} have been reported in the realization involving superconducting radio frequency cavities together with drum oscillators. A major motivation  is to study the fundamentals of quantum mechanics.

\begin{figure*}
\centering
\includegraphics[width=\textwidth]{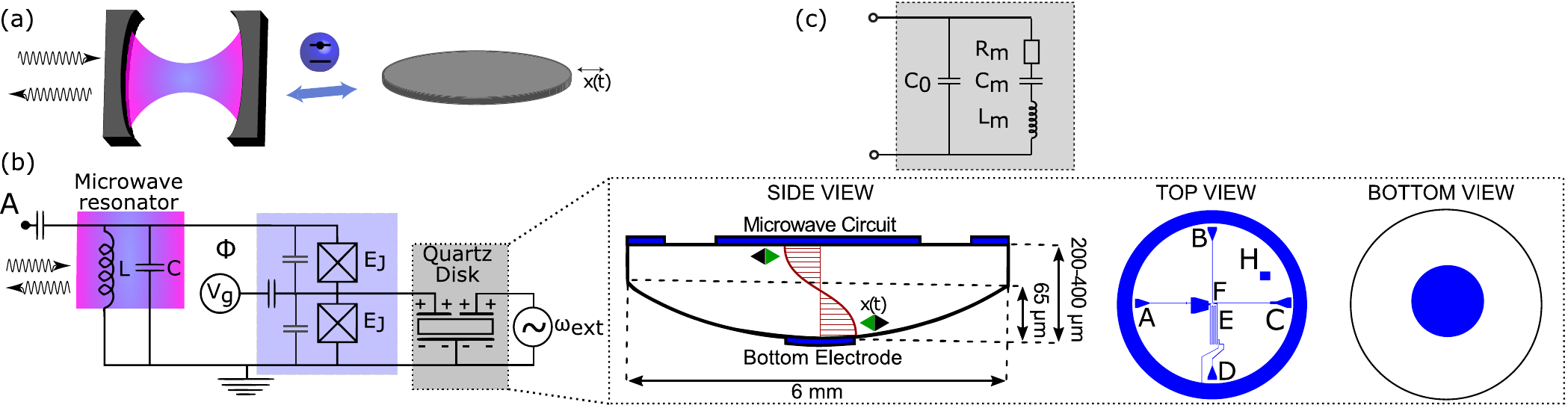}
  \caption{\textit{Device schematics.} (a) Illustration of a quantum two-level system (qubit) mediating an optomechanical interaction between a macroscopic quartz oscillator and an electromagnetic cavity. (b) Representation of the quartz disk oscillator, qubit, and microwave resonator. The charge qubit is contained within the blue box. The circuit is fabricated directly on top of the flat surface of a quartz disk with plano-convex profile, which traps the mechanical energy in the middle, thereby increasing mechanical Q-values \cite{onoe2005, dataint1998, hirama1997, cumpson1990}. The lowest shear mode at \SI{7}{\mega\hertz} couples to the microwave circuit. Blue denotes aluminum metallization, and white is bare quartz. The bonding pads are labeled as: A is the input/output for the reflection measurement, B is the gate bias, C is a test junction, D is the flux bias, E is the on-chip inductor $L,$ F is the junctions and island region, and H is used for actuating the quartz vibrations. (c) Equivalent electrical resonator of the piezoelectric quartz oscillator.}
  \label{fig:system}
\end{figure*}

Here we propose and demonstrate a new cavity optomechanical scheme which involves a genuinely macroscopic mechanical oscillator near the quantum limit (see Fig.~\ref{fig:system}). We use a 6 mm diameter, 20 mg quartz disk oscillator, whose vibrations at the lowest shear mode frequency $\omega_m/2 \pi \simeq 7$ MHz we observe at a temperature of 35 mK. The quartz disk is used both as the mechanical oscillator, and as a substrate for fabricating the superconducting micro circuit. The latter includes a charge qubit and a transmission line resonator, which together form a single, effective cavity. The energy of the qubit depends on the charge in its vicinity, and the frequency of the effective cavity thus becomes charge-sensitive \cite{sillanpaa2004, heikkila2014, pirkkalainen2015_2}. Due to the piezoelectricity of the quartz oscillator, the deformation corresponding to the mechanical vibrations induces charge on the chip surface. Hence we obtain a cavity optomechanical setup, where electromagnetic fields and mechanical motion interact in a confined volume \cite{aspelmeyer2014}, as illustrated in Fig.~\ref{fig:system}a. We repeat that the underlying coupling mechanism is not the usual parametric coupling via a movable capacitance, but the qubit converts a linear coupling into an effective parametric coupling. A related theory proposal \cite{woolley2016} was presented recently.

Let us study a piezoelectric disk having thickness $t$, shear mode stress coefficient $e_s$ of the material,  shear modulus $Y_s$, and relative permittivity $\epsilon_r$. One further defines the dimensionless piezoelectric coupling coefficient $K_0^2=e_s^2/(\epsilon_r \epsilon_0 Y_s)$. A shear deformation by a characteristic distance $x$ corresponds to a shear strain $\lambda_s=x/t$, and generates a piezoelectric surface charge density $\sigma_q=\lambda_s e_s$. A piezoelectric oscillator is made by metallizing both surfaces of the chip over an area $A$. The geometric capacitance in the plate-capacitor approximation is then $C_0 =\epsilon_r \epsilon_0 A/t$. The oscillator can be represented as an equivalent series LCR resonator (Fig. \ref{fig:system}c) with the effective parameters $C_m=K_0^2 C_0$, $L_m=[(2 \pi \omega_m)^2 C_m]^{-1}$ and $R_m=(\omega_m C_m Q)^{-1}$, with $Q$ the mechanical quality factor. The corresponding quantized harmonic oscillator exhibits zero-point vibrations of an amplitude $x_{\m{zp}} =\sqrt{\hbar / 2 M \omega_m}$, where $M$ is the effective mass. In case of our macroscopic oscillator, $x_{\m{zp}}$ is very small, in the range of \SIrange{e-18}{e-19}{\meter}.

We continue by discussing the interaction of the piezo motion with the qubit. The charge qubit consists of two small-area Josephson junctions defining an island. The junctions are supposed to have equal Josephson energies $E_J$. The junctions, island, and a capacitive gate contribute to the sum capacitance $C_{\Sigma}$ of the qubit, which provides the charging energy of a single electron as $E_C=e^2/2C_{\Sigma}$. The charge qubit limit, $E_J/E_C \lesssim 1$, entails a strong dependence of the qubit energy on the charge $q_g$ polarized on the qubit island. In the following, we use a dimensionless charge $n_g=q_g/2e$ in units of the Cooper-pair charge $2e$. Besides the gate charge, the qubit energy can be controlled by a magnetic flux $\Phi$ in the superconducting loop shown in Fig.~\ref{fig:system}b. The Hamiltonian of the qubit is $H_q = B_x /2\sigma_x - B_z/2\sigma_z$. The effective magnetic fields are $B_x=2 E_J \cos (\pi \Phi/\Phi_0)$, $B_z=4 E_C (1-2n_g)$, and $\sigma_x$ and $\sigma_z$ are the Pauli spin matrices. Using the equivalent circuit in Fig.~\ref{fig:system}c, the qubit-piezo interaction can be derived as $H_{qm}=g_m \sigma_z (b^\dagger - b)$, where $b^\dagger$, $b$ are operators of the oscillator. The coupling energy is $g_m=\frac{2 E_C}{e} \sqrt{\frac{C_m \hbar \omega_m}{2}}$.

\begin{figure*}
\centering
\includegraphics[width=\textwidth]{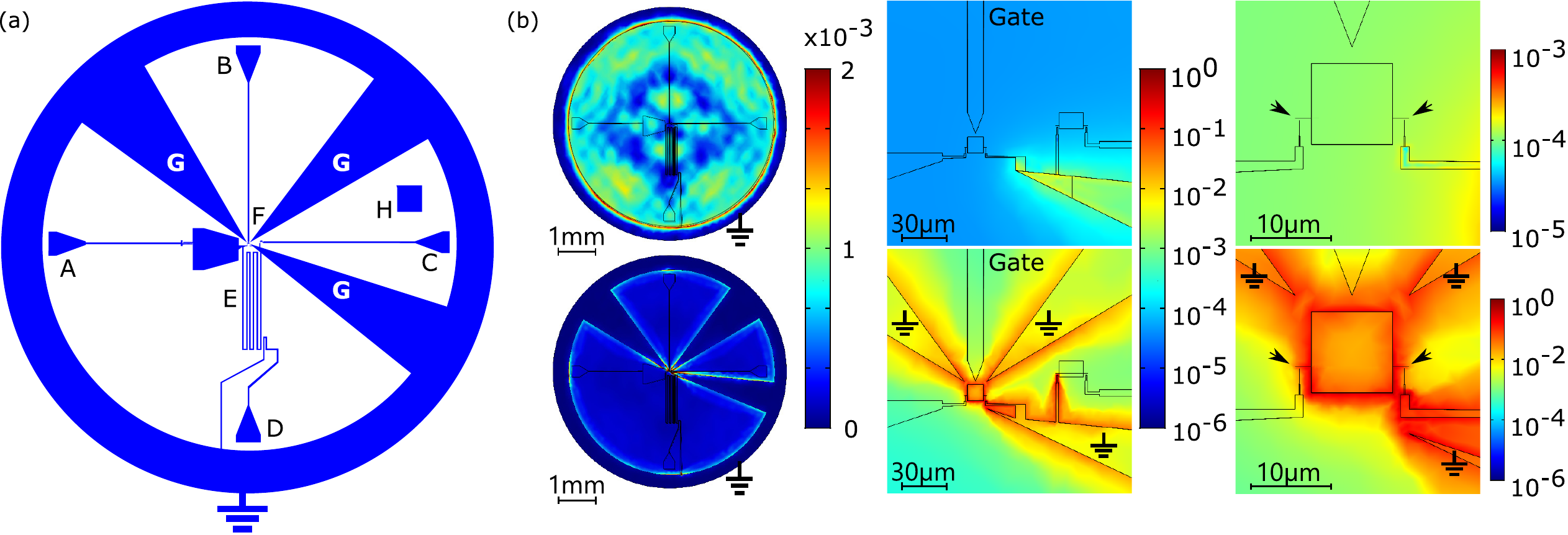}
  \caption{\textit{Focusing of the piezo charge.} (a) Circuit layout involving three grounded spikes (labeled G). Other labels are as in \fref{fig:system}b. Blue denotes aluminum metallization, and white is bare quartz. (b) Simulation of the piezoelectric surface charge density induced by the lowest mechanical shear mode vibrations of the quartz disk. The top row shows the results for the layout in Fig.~\ref{fig:system}b, while the bottom row displays the behavior of a charge focusing layout shown in (a). The values are scaled by the maximum in the focusing layout. The qubit junctions are marked by arrows.}
  \label{fig:sim}
\end{figure*}

\begin{figure}
  \centering
    \includegraphics[width=\columnwidth]{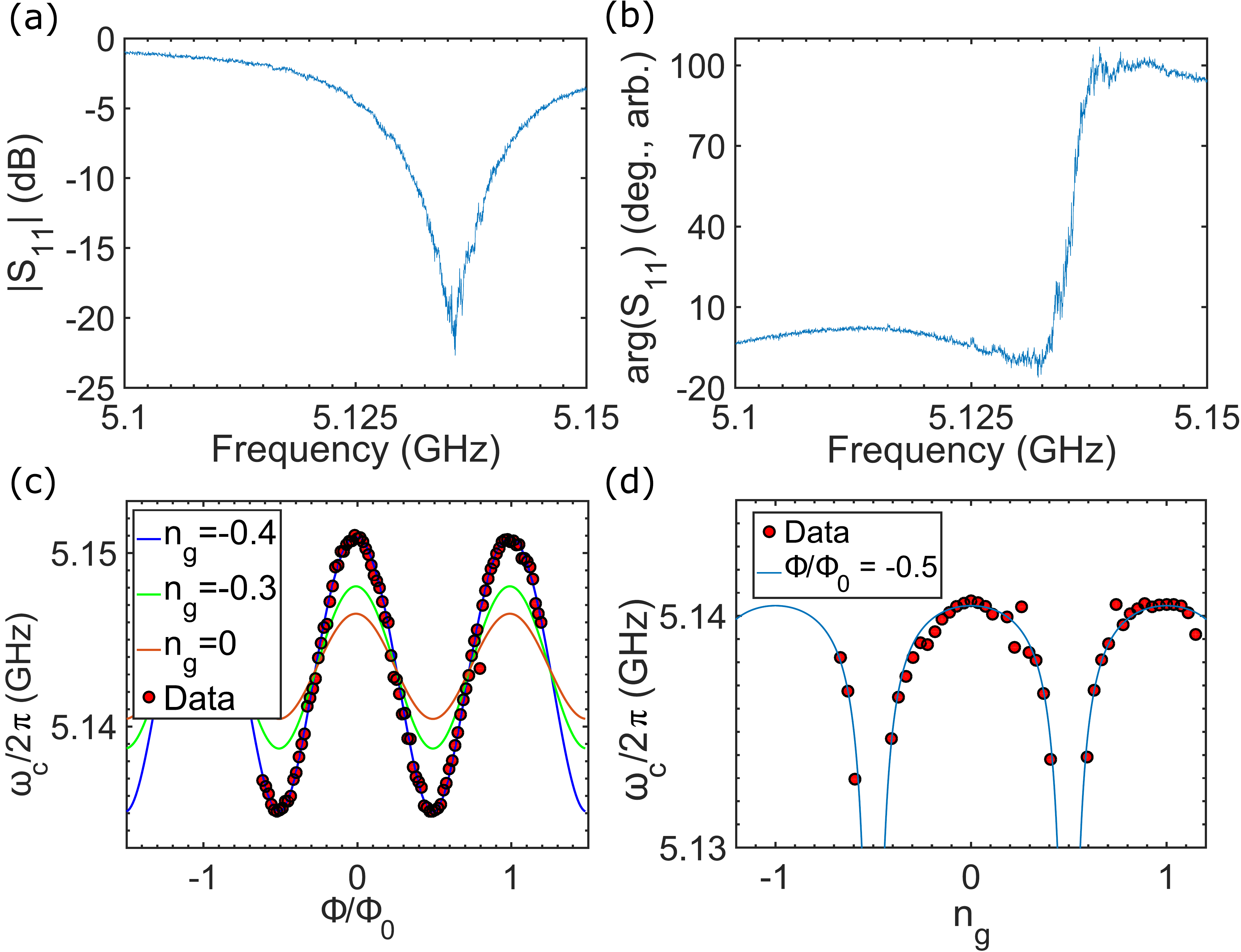}
  \caption{\textit{Response of the charge-sensitive microwave cavity.} (a) Magnitude and (b) Phase of the microwave reflection. (c) and (d) represent the frequency shift of the cavity resonance due to magnetic flux $\Phi$ in the superconducting loop, and the gate charge $n_g$, respectively. The dots depict the data when the device is biased with (c) $n_g\approx 0.4$ and (d) $\Phi/\Phi_0\approx 0.5$. The solid curves correspond to the theoretical model assuming the parameters average $E_J=\SI{0.12}{\kelvin}$, $E_J/E_C=\num{0.45}$, junction resistance asymmetry $d=\num{0.25}$, $L=\SI{2.94}{\nano\henry}$, and $C=\SI{325.5}{\femto\farad}$.}
  \label{fig:data_set}
\end{figure}

\begin{figure}
  \centering
    \includegraphics[width=0.85 \columnwidth]{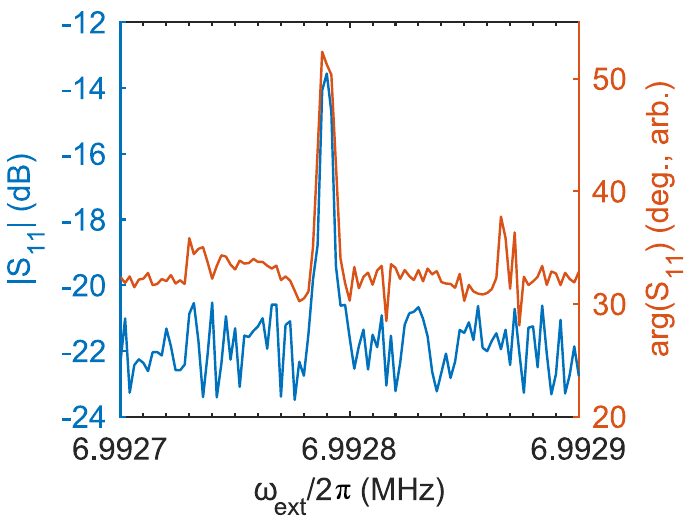}
  \caption{\textit{Driven response of the quartz oscillator.} The reflection $S_{11}$ of a weak probe tone of a fixed frequency $\sim \omega_c$, while the mechanical vibrations are excited with a voltage of $\SI{1}{\micro\volt}$ at a varying frequency $\omega_{\m{ext}}$.}
  \label{fig:driven}
\end{figure}

\begin{figure}
  \centering
			    \includegraphics[width= 0.9 \columnwidth]{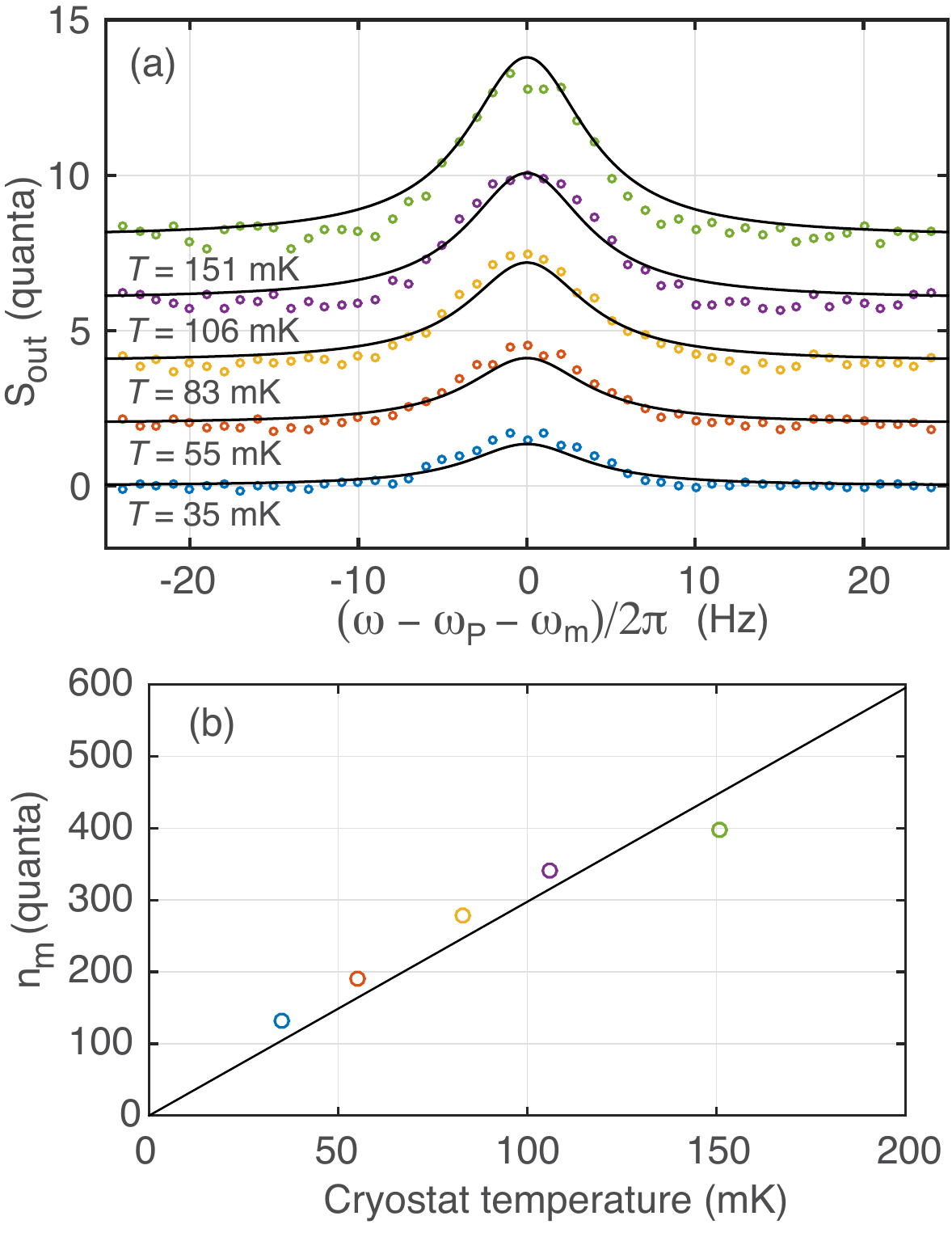}
  \caption{\textit{Thermal motion near the quantum limit.} (a) Spectral density showing the motional sideband at different cryostat temperatures. The circles are the experimental data, and the solid lines represent the theoretical model. The pump frequency was $\omega_p\simeq\omega_c+\omega_m$. The curves are shifted vertically by 2 units for clarity. (b) Phonon number as a function of the cryostat temperature. The dots depict the area below Lorentzian fits to the data in (a), and the solid line represents a linear fit crossing the origin.}
  \label{fig:sideb}
\end{figure}

In order to create the optomechanical interaction, we include in the circuit design a transmission line resonator, which shunts the qubit, as presented in Fig.~\ref{fig:system}b. We label the qubit-cavity coupling energy as $g_{qc}$. The radiation-pressure coupling can be derived by considering the microwave cavity having the Hamiltonian $H=\omega_c a^\dagger a$. The cavity frequency  $\omega_c$   is Lamb shifted from the intrinsic value $\omega_{c0}$ as $\omega_c\simeq \omega_{c0}+ g_{qc}^2/\Delta$, where $\Delta$ is the detuning between qubit and cavity frequencies. In the presence of the piezo coupling, and supposing the qubit stays in the ground state, the detuning becomes $\Delta\Rightarrow\Delta-g_m(b^\dagger-b)$. The cavity Hamiltonian now assumes the form $H=g_0a^\dagger a(b^\dagger+b)$, where the (single photon) radiation-pressure coupling is $g_0=(g_{qc}/\Delta)^2 g_m$. The analysis above holds only if $\Delta\gg g_{qc}$, so it is not immediately clear how large a ratio $g_0/g_m$ one can achieve.

With typical parameters \cite{heikkila2014, pirkkalainen2015_2}, $g_{qc}/\omega_c\approx 0.1$, thus the qubit and cavity are close to the ultra-strong coupling regime \cite{niemczyk2010}, and they had better be considered as a single effective cavity. In order to make quantitative estimates of the radiation-pressure coupling, we treat the effective cavity as a charge-tunable resonator. The radiation pressure coupling is $g_0\equiv\frac{\partial\omega_c}{\partial x}x_{\m{zp}}=\frac{\partial\omega_c}{\partial n_g} n_{\m{zp}}^0$, where we define the number of charges in the surface area under the qubit island corresponding to zero-point motion, $n_{\m{zp}}^0=\frac{\partial n_g}{\partial x} x_{\m{zp}}$.
In terms of the piezo-induced charge density $\sigma_q$, the number of charges in the surface area under the qubit island with area $A_{qb}$ is $n_{\m{zp}}^0=\sigma_q A_{qb}$.

Next we will work out some numbers using the parameters of the experiment discussed below. We now suppose the piezo charge density is uniformly spread across the quartz disk surface, thus $n_{\m{zp}}^0=\frac{x_{\m{zp}} e_s A_{qb}}{2 e t}$. Quartz has the material parameters $e_s\approx \SI{0.1}{\coulomb\per\square\meter}$, $K_0^2\approx 0.01$, $Y_s\approx\SI{30}{\giga\pascal}$ and $\epsilon_r\approx 4.0$. We use circular disks with surface area $A\approx\pi \cdot (\SI{3}{\milli\meter})^2$ and thickness $t\approx \SI{400}{\micro\meter}$. Given a typical qubit island $A_{qb}\approx( \SI{10}{\micro\meter})^2$, we obtain $n_{\m{zp}}^0\approx\num{e-8}$, which is small compared to typical charge sensitivities in microwave single-electron devices, in the range of \numrange{e-6}{e-4}$/\sqrt{\m{Hz}}$. With a good choice of parameters of the qubit and the microwave cavity \cite{pirkkalainen2015_2}, the modulation of cavity frequency as a function of charge is of the order $\frac{\partial\omega_c}{\partial n_g}\approx (2\pi) \cdot \SI{50}{\mega\hertz}$, and we estimate $g_0 \approx 2\pi\cdot 1$~Hz. This is two orders of magnitude smaller than what is obtained in typical aluminum drum resonators \cite{teufel2011}, and is mostly limited by the small island $A_{qb}/A \ll 1$.



Instead of enlarging the qubit island in order to increase $g_0\propto n_{\m{zp}}^0=\sigma_q A_{qb}$, our primary strategy is to focus most of the piezo charge nearby the island. This is achieved through manipulation of the mechanical mode shape, thereby locally increasing $\sigma_q$. We use Comsol Multiphysics simulations  to find the charge density which arises when the lowest flexural mode has a given energy. The basic circuit layout used in earlier related work \cite{pirkkalainen2015_2} is shown in Fig.~\ref{fig:system}b, and the corresponding normalized charge density at various levels of zoom in the top row of Fig.~ \ref{fig:sim}b. As seen in Fig.~\ref{fig:sim}b, the charge density is approximately uniformly spread all over the chip. We suppose this roughly corresponds to the uniform distribution discussed above. For the final design shown in Fig.~\ref{fig:sim}a, we add grounded spikes which extend from the ground planes surrounding the chip towards the qubit island. The spikes strongly enhance the nearby electric field and act as anchors, focusing most of the piezoelectric strain / charge to the center of the disk (bottom row of Fig.~\ref{fig:sim}b). Comparing the realizations in the top and bottom rows of Fig.~\ref{fig:sim}b, the difference in the integrated charge around the qubit island differs by two orders of magnitude in favor of the charge focusing design, translating into a dramatic increase in the radiation pressure coupling. 
We hence expect $g_0/2\pi \approx\SI{150}{\hertz}$, on par with mesoscopic aluminum drum oscillators.

The measurements were carried out in a dilution refrigerator with a base temperature of \SI{35}{\milli\kelvin}. The microwave tones are applied to the port A of Fig.~\ref{fig:system}, and we record the scattered signal from the same port. The cavity linewidth below \SI{200}{\milli\kelvin} was  $\kappa/2\pi\approx\SI{7}{\mega\hertz}$.
In Fig. \ref{fig:data_set}a and Fig. \ref{fig:data_set}b we show examples of the cavity resonance absorption and phase shift. The cavity frequency is sensitive both to  flux and charge as seen in Fig. \ref{fig:data_set}c and Fig. \ref{fig:data_set}d, respectively. The solid lines represent a theoretical fit produced by numerical diagonalization of the coupled qubit - resonator Hamiltonian, displaying an excellent agreement.

We now turn the discussion on observing the quartz vibrations by cavity optomechanical means. We first studied the driven motion by exciting the quartz through the actuation pad (labeled H in Fig.~\ref{fig:system}), and probed possible nonlinear changes under an intense driving. A piezo charge which is a substantial fraction of one electron will change the time averaged cavity response (see Fig.~\ref{fig:data_set}d), causing a clear signature in the $S_{11}$ response to a weak probe tone. Figure \ref{fig:driven} shows this type of the detection, revealing the lowest shear mode resonance at the expected frequency. From the simulations, we  obtain an independent estimate of the piezo charge under the experimental driving conditions. The simulated value in the range of $\sim 0.1 \, e$ is in a reasonable agreement with the observation.

The benchmark of cavity optomechanics is the measurement of the motional sidebands due to thermal vibrations. Here, a pump tone $\omega_p$ is applied near the cavity resonance frequency, and the motional sidebands appear in the spectrum at the frequencies $\omega_p \pm\omega_m$. The relevant coupling figure of merit is the effective coupling $G=g_0\sqrt{n_P}$, where $n_P$ is the photon number arising in the cavity via pumping. Our cavities reach a maximum $n_P\approx$ \numrange{e1}{e2} limited by the Josephson nonlinearity \cite{sillanpaa2004, pirkkalainen2015_2}. In Fig.~\ref{fig:sideb}a we show the results of such measurement for different cryostat temperatures. The pump power is optimized for maximizing the signal. At fixed pumping conditions, the area under the power spectrum is expected to be proportional to the mode temperature, since the energy of a thermally actuated oscillator will be $\simeq k_B T$. Indeed, as seen in Fig.~\ref{fig:sideb}a, the peaks grow when the temperature is increased. From fits to the theory \cite{rocheleau2010}, we extracted mechanical Q-value of \num{6.4e5}, $g_0\approx\SI{180}{\hertz}$ with a maximum $n_P\approx17$, agreeing with the simulated $g_0$ for our charge focusing design. We follow the usual practice and calibrate the equilibrium phonon number $n_m^T$ by relying on the expected linear temperature dependence $n_m^T\approx k_B T / \hbar\omega_m$. Based on the data and a linear fit shown in Fig. \ref{fig:sideb}b, we conclude that the shear mode thermalizes down to roughly  $\approx\SI{40}{\milli\kelvin}$, which corresponds to only $n_m \approx 130$ phonons in the mm-sized vibrating disk.

Apart from observing the thermal vibrations, cavity optomechanical techniques can allow for controlling them. The basic phenomenon is the optically induced change in damping from the intrinsic value $\gamma$, $\gamma_{\m{opt}}=\pm 4 G^2/\kappa$, leading either to sideband cooling or lasing. The final phonon number under sideband cooling is $n_m=n_m^T\gamma/(\gamma + \gamma_{\m{opt}})$. In the present experiment, cooling of only a few percent can be expected, which is too small to be directly measurable.

In order to foresee an optimized device, we keep the microwave circuit and qubit junctions unchanged. The optimal island size is a balance between the piezo charge sensed by the qubit, and increased qubit capacitance which reduces the charge dispersion. Choosing a roughly $\sim (380 \; \si{\micro\meter})^2$ sized island with proper charge focusing, we simulate that the charge  is further enhanced by 3 orders of magnitude. The coupling is strongly enhanced up to $g_0/2\pi\sim\SI{8}{\kilo\hertz}$. With the current mechanical Q, larger coupling will allow for sideband cooling down to basically the ground state, $n_m\sim 1$  quanta. With macroscopic quartz oscillators, high Q values $> 10^8$ have been obtained \cite{Besson, goryachev2012}. Such Q would allow us to cool the improved device deep in the ground state. Besides the effective cavity, the piezo motion interacts with the qubit-like mode in the microwave circuit. In the limit of high $E_J/E_C\gg1$, the qubit-motion coupling energy becomes $g'_m=\sqrt{\omega_m\omega_{01}}\sqrt{\frac{C_m}{C_\Sigma}}$, where $\omega_{01}$ is the qubit frequency. With an island size of the order $\sim (1 \; \m{mm})^2$, without charge focusing, one obtains that $C_0\approx\SI{100}{\femto\farad}$, $C_m\approx\SI{1}{\femto\farad}$, and $g'_m/2\pi\approx\SI{5}{\mega\hertz}$. This value is similar to the coupling obtained with micron-sized oscillators, offering prospects to delicate quantum control via the qubit \cite{etaki2008, lahaye2009, pirkkalainen2013, gustafsson2014, lecocq2015_2}.

To conclude, we demonstrated cavity optomechanics on a truly macroscopic mechanical oscillator near the quantum limit. Focusing of piezo charge allowed coupling of the mechanical oscillations of a mm-sized quartz disk, via a charge qubit, to a microwave cavity. Future challenges include observing the back-action of the cavity, or, in other words, back-action of a single Cooper pair on a macroscopic moving object. We foresee that this is feasible with realistic parameters, and we can reach the quantum ground state in the near future. 

\begin{acknowledgments}
We thank Francesco Massel  for useful discussions. This work was supported by the Academy of Finland (contract 250280, CoE LTQ, 275245), the European Research Council (615755-CAVITYQPD), and by the Finnish Cultural Foundation. The work benefited from the facilities at the OtaNano - Micronova Nanofabrication Center and at the Low Temperature Laboratory. 

\end{acknowledgments}

\bibliography{PaperI}

\end{document}